\documentclass[iop]{emulateapj}
\usepackage{epsfig}
\usepackage{ifthen}
\usepackage{lipsum}
\usepackage{graphicx}

\newcommand{\kms}{\ensuremath{\rm km\,s^{-1}}}
\newcommand{\ms}{\ensuremath{\rm m\,s^{-1}}}
\newcommand{\msday}{\ensuremath{\rm m\,s^{-1}\,day^{-1}}}
\newcommand{\logg}{\ensuremath{\log{g}}}
\newcommand{\teff}{\ensuremath{T_{\rm eff}}}
\newcommand{\vsini}{\ensuremath{v \sin{i}}}
\newcommand{\feh}{\rm{[Fe/H]}}
\newcommand{\meh}{\rm{[m/H]}}
\newcommand{\rsun}{\ensuremath{R_\sun}}
\newcommand{\msun}{\ensuremath{M_\sun}}
\newcommand{\rstar}{\ensuremath{R_\star}}
\newcommand{\mstar}{\ensuremath{M_\star}}

\newcommand{\rplrs}{\ensuremath{R_{\rm P}/R_\star}}
\newcommand{\rsa}{\ensuremath{R_\star/a}}

\newcommand{\rjup}{\ensuremath{R_{\rm J}}}
\newcommand{\mjup}{\ensuremath{M_{\rm J}}}
\newcommand{\vt}{V_{\rm T}}
\newcommand{\bt}{B_{\rm T}}
\newcommand{\bjdtdb}{\ensuremath{\rm {BJD_{TDB}}}}

\newcommand{\lsun}{\ensuremath{\,L_\Sun}}
\newcommand{\mj}{\ensuremath{\,M_{\rm J}}}
\newcommand{\rj}{\ensuremath{\,R_{\rm J}}}
\newcommand{\fave}{\langle F \rangle}
\newcommand{\fluxcgs}{10$^9$ erg s$^{-1}$ cm$^{-2}$}

\newcommand{\twomass}{2MASS~05131092+3319054}                  
\newcommand{\gsc}{GSC~2393-00852}                              
\newcommand{\tycho}{TYC~2393-852-1}                            
\newcommand{\bd}{BD~+33 977}                                   
\newcommand{\hd}{HD~33643}                                     
\newcommand{\vmag}{\ensuremath{8.54}}                          
\newcommand{\vtmag}{\ensuremath{8.612}}                         
\newcommand{\btmag}{\ensuremath{9.074}}                        
\newcommand{\mstarmulti}{\ensuremath{1.535}}                   
\newcommand{\rstarmulti}{\ensuremath{1.732}}                   
\newcommand{\loggmulti}{\ensuremath{4.149}}                    
\newcommand{\loggspc}{\ensuremath{4.23}}                      
\newcommand{\teffmulti}{\ensuremath{6789}}                     
\newcommand{\teffspc}{\ensuremath{6779}}                       
\newcommand{\fehmulti}{\ensuremath{0.139}}                    
\newcommand{\fehspc}{\ensuremath{0.12}}                      
\newcommand{\vsinispc}{\ensuremath{73}}                        
\newcommand{\period}{\ensuremath{2.7347749}}                   
\newcommand{\rplmulti}{\ensuremath{1.533}}                     
\newcommand{\mplmulti}{\ensuremath{1.28}}                      
\newcommand{\depthmmag}{\ensuremath{8.28}}                        

\begin{document}


\shorttitle{Kelt-7b}
\title{KELT-7\MakeLowercase{b}: A hot Jupiter transiting a bright V=$\vmag$ rapidly rotating F-star}


\shortauthors{Bieryla et al}
\author{Allyson Bieryla\altaffilmark{1}, Karen Collins\altaffilmark{2}, Thomas G. Beatty\altaffilmark{3}, Jason Eastman\altaffilmark{1,4}, 
Robert J. Siverd\altaffilmark{4}, Joshua Pepper\altaffilmark{6}, B. Scott Gaudi\altaffilmark{7}, Keivan G. Stassun\altaffilmark{5,8}, 
Caleb Ca\~nas\altaffilmark{1}, David W. Latham\altaffilmark{1}, Lars A. Buchhave\altaffilmark{1,9}, Roberto Sanchis-Ojeda\altaffilmark{10,11}, 
Joshua N. Winn\altaffilmark{12}, Eric L. N. Jensen\altaffilmark{13}, John F. Kielkopf\altaffilmark{2}, Kim K. McLeod\altaffilmark{14}, 
Joao Gregorio\altaffilmark{15}, Knicole D. Col\'on\altaffilmark{6}, Rachel Street\altaffilmark{4}, Rachel Ross\altaffilmark{4}, 
Matthew Penny\altaffilmark{7}, Samuel N. Mellon\altaffilmark{16}, Thomas E. Oberst\altaffilmark{16}, Benjamin J. Fulton\altaffilmark{17,18},
Ji Wang\altaffilmark{19}, Perry Berlind\altaffilmark{1}, Michael L. Calkins\altaffilmark{1}, Gilbert A. Esquerdo\altaffilmark{1}, 
Darren L. DePoy\altaffilmark{20}, Andrew Gould\altaffilmark{7}, Jennifer Marshall\altaffilmark{20}, Richard Pogge\altaffilmark{7}, 
Mark Trueblood\altaffilmark{21}, Patricia Trueblood\altaffilmark{21}}


\altaffiltext{1}{Harvard-Smithsonian Center for Astrophysics, Cambridge, MA 02138 USA; email: abieryla@cfa.harvard.edu}

\altaffiltext{2}{Department of Physics and Astronomy, University of Louisville, Louisville, KY 40292, USA}

\altaffiltext{3}{Department of Astronomy and Astrophysics, Pennsylvania State University, University Park, PA 16801, USA}

\altaffiltext{4}{Las Cumbres Observatory Global Telescope Network, 6740 Cortona Drive, Suite 102, Santa Barbara, CA 93117, USA}

\altaffiltext{5}{Department of Physics and Astronomy, Vanderbilt University, Nashville, TN 37235, USA}

\altaffiltext{6}{Department of Physics, Lehigh University, Bethlehem, PA 18015, USA}

\altaffiltext{7}{Department of Astronomy, The Ohio State University, 140 W. 18th Ave., Columbus, OH 43210, USA}

\altaffiltext{8}{Department of Physics, Fisk University, Nashville, TN 37208, USA}

\altaffiltext{9}{Niels Bohr Institute, University of Copenhagen, DK-2100, Denmark, and Centre for Star and Planet Formation, 
Natural History Museum of Denmark, DK-1350 Copenhagen}

\altaffiltext{10}{Department of Astronomy, University of California Berkeley, Berkeley, CA 94720, USA}

\altaffiltext{11}{NASA Sagan Fellow}

\altaffiltext{12}{Department of Physics and Kavli Institute for Astrophysics and Space Research, Massachusetts Institute of 
Technology, Cambridge, MA 02139, USA}

\altaffiltext{13}{Department of Physics and Astronomy, Swarthmore College, Swarthmore, PA 19081, USA}

\altaffiltext{14}{Wellesley College, Wellesley, MA 02481, USA}

\altaffiltext{15}{Atalaia Group and Crow-Observatory, Portalegre, Portugal}

\altaffiltext{16}{Westminster College, New Wilmington, PA 16172, USA}

\altaffiltext{17}{Institute of Astronomy, University of Hawaii at Manoa, 2680 Woodlawn Drive, Honolulu, HI 96822, USA}

\altaffiltext{18}{National Science Foundation Graduate Research Fellow}

\altaffiltext{19}{Yale University, New Haven, CT 06520, USA}

\altaffiltext{20}{George P. and Cynthia Woods Mitchell Institute for Fundamental Physics and Astronomy, Texas A and M University, 
College Station, TX 77843-4242, USA}

\altaffiltext{21}{Winer Observaotry, Sonoita, AZ 85637, USA}


\begin{abstract}
We report the discovery of KELT-7b, a transiting hot Jupiter with a mass of $\mplmulti \pm 0.18$ \mjup, radius of 
$\rplmulti_{-0.047}^{+0.046}$\,\rjup, and an orbital period of $\period \pm 0.0000039$\ days.
The bright host star (\hd; KELT-7) is an F-star with $V=\vmag$, \teff $=\teffmulti_{-49}^{+50}$ K, \feh $=\fehmulti_{-0.081}^{+0.075}$,
 and \logg $=\loggmulti \pm 0.019$.
It has a mass of $\mstarmulti_{-0.054}^{+0.066}$\,\msun, a radius of $\rstarmulti_{-0.045}^{+0.043}$\,\rsun, and is the fifth most massive, 
fifth hottest, and the 
ninth brightest star known to host a transiting planet. It is also the brightest star around which KELT has discovered a  
transiting planet. Thus, KELT-7b
is an ideal target for detailed characterization given its relatively low surface gravity, high equilibrium temperature, and bright
host star. The rapid rotation of the star ($\vsinispc \pm 0.5$\,\kms) results in a Rossiter-McLaughlin effect with an unusually large
amplitude of several hundred \ms. We find that the orbit normal of the planet is likely to be well-aligned with the stellar spin axis,
with a projected spin-orbit alignment of $\lambda=9.7 \pm 5.2$ degrees. This is currently the second most rapidly rotating star to have
a reflex signal (and thus mass determination) due to a planetary companion measured.

\end{abstract}

\keywords{
   planetary systems ---
   stars: individual (KELT)
   techniques: spectroscopic, photometric
}


\section{Introduction}
\label{sec:intro}

Transiting planets that orbit bright host stars are of great value to the exoplanet community. Bright host 
stars are ideal candidates for follow-up because the higher photon flux generally allows for a wider array 
of follow-up observations, more precise determination of physical parameters, and better ability to diagnose
and control systematic errors. As a result, bright transiting systems have proven to be important laboratories
for studying atmospheric properties of the planets through transmission and emission spectroscopy, for measuring
the spin-orbit alignment of the planet orbits, and for determining precise stellar parameters (see \citet{winn:2011} 
for a review).

\begin{figure}[]
\plotone{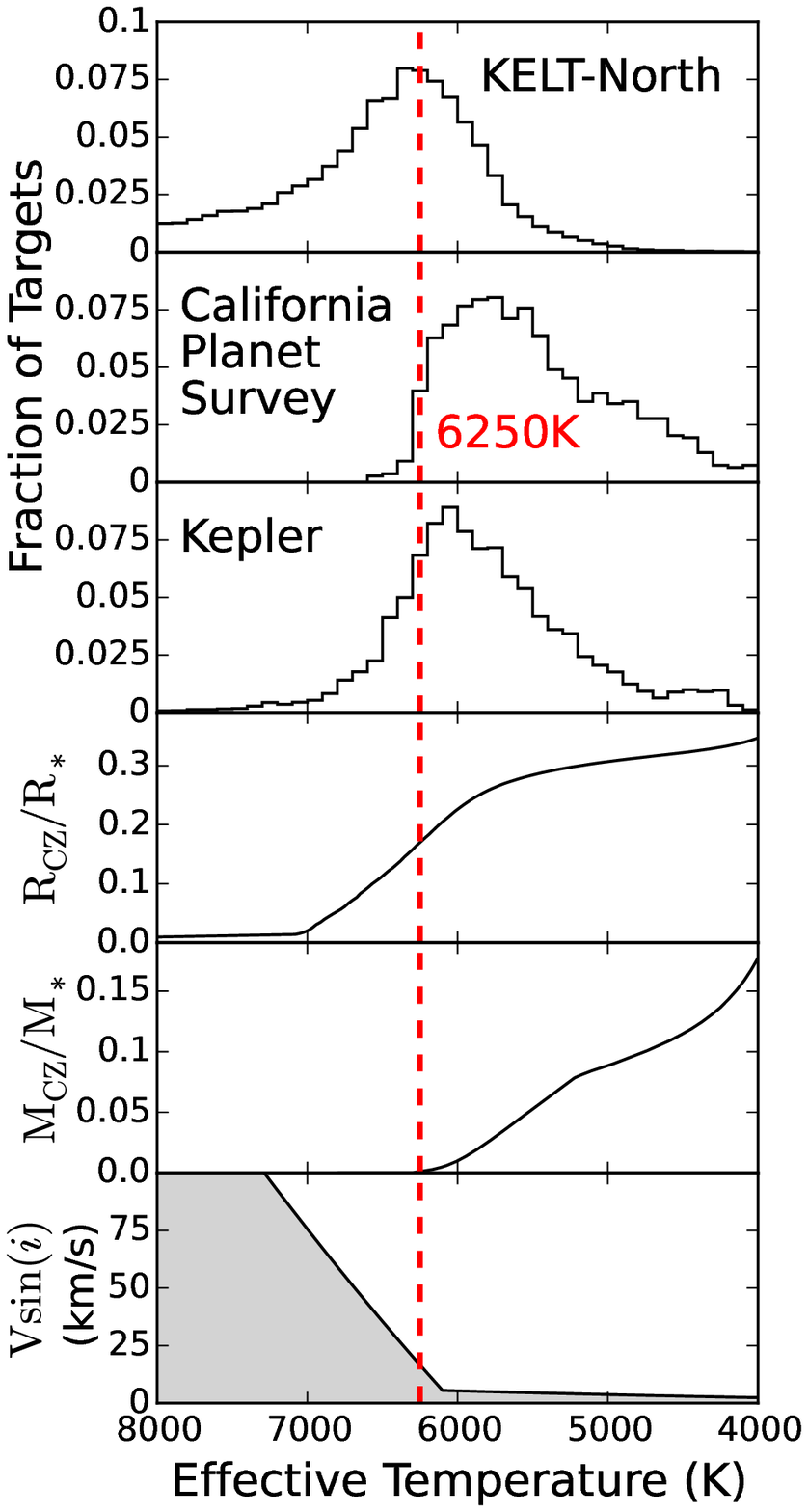}
\caption[]{
The top three panels show the effective temperature distribution for
the stars targeted by the KELT-North transit survey \citep{siverd:2012}, the California
Planet Survey (CPS) radial velocity (RV) search \citep{wright:2004}, and the Kepler
mission (stars observed for all $16$ quarters and with $\logg\ > 4.0$ according to
Kepler Q1-Q16 Stellar Parameters Database 
\footnote{http://exoplanetarchive.ipac.caltech.edu, NASA Exoplanet Archive}).
The fourth and fifth panels from the top show the relative
depth and mass of outer stellar convective zones at these temperatures
\citep{vansaders:2012}, while the sixth, bottom, panel shows the
observed stellar \vsini\ distribution \citep{reiners:2003}. The red
dashed line at $6250$K shows the approximate location of the Kraft Break
\citep{kraft:1970}.
\label{fig:teffhist}}
\end{figure}

The Kilodegree Extremely Little Telescope (KELT) transit survey
\citep{pepper:2007} was designed to detect transiting planets around
bright ($8 < V < 10$) stars. Very few ($\sim 3\%$) of the known
transiting planet host stars are in this brightness range. This is because
this range spans the gap between radial velocity surveys on the bright
end, and the saturation limit of the 
majority of ground-based transit surveys on the faint
end.  The KELT-North (KELT-N) telescope targets this range using a
small-aperture ($42$ mm) camera with a a wide field of view of $26^{\circ} \times 26^{\circ}$.
It observes 13 fields at declination of 31.7 degrees, roughly equally
spaced in right ascension, in total covering approximately $40\%$ of the
Northern sky. The KELT-N survey has been in operation since 2006 and
candidates have been actively vetted since April 2011.

The KELT-N survey has already announced four
planet discoveries. KELT-1b \citep{siverd:2012} is a
$27 \mjup$ brown dwarf transiting a $V = 10.7$ F-star. 
KELT-2Ab \citep{beatty:2012} is a hot Jupiter transiting the
bright ($V = 8.77$) primary star in a visual binary system.
KELT-3b \citep{pepper:2013} is a hot Jupiter transiting
a $V = 9.8$ slightly evolved late F-star. KELT-6b \citep{collins:2014}
is a mildly-inflated Saturn-mass planet transiting
a metal poor, slightly evolved late F-star. 

Because of its brighter magnitude range, the sample of host stars
surveyed by KELT has a higher percentage of luminous stars than most
transit surveys.  This
luminous subsample includes giants, as well as hot main-sequence stars
and subgiants.  Indeed, all five of the KELT-N discoveries to date (including
KELT-7b) orbit F stars with $\teff>6100$K.  Such hot stars are typically avoided by
radial velocity surveys.  There is a transition between slowly and rapidly rotating
stars known as the Kraft break \citep{kraft:1970, kraft:1967}. Stars hotter than the 
Kraft break around $\teff=6250$K typically have higher rotation velocities, making 
precision radial velocities more difficult. These higher rotation velocities reflect
the angular momentum from formation, which is conserved as the stars
evolve due to the lack of convective envelope.  The lack of a
convective envelope results in weak magnetic fields and ineffective
magnetic breaking from stellar winds (e.g., \citet{vansaders:2013}).  

These points are illustrated in Figure \ref{fig:teffhist}, which shows the distribution
of effective temperatures for the KELT-N stellar sample, the sample of
stars targeted by Kepler, and a representative radial velocity
survey. The KELT-N targets plotted
here are all of the bright ($V<11$) putative dwarf stars in the survey
which were selected by a reduced proper motion cut and with
temperatures calculated from their J-K colors. 
Approximately $40,000$ KELT-North targets ($55\%$) are hotter than
$6250$K, $28$ of the CPS targets ($2.3\%$) are hotter than $6250$K, and approximately
$20,000$ of the Kepler targets ($20\%$) are hotter than $6250$K. Also shown are theoretical estimates of the mass and radius of
the convective envelope as a function of \teff\ for stars with solar
metallicity and an age of $1$ Gyr \citep{vansaders:2012}, as well as the upper envelope
of observed rotation velocities as a function of \teff\ based from \citet{reiners:2003}.

Hot stars pose both opportunities and challenges for transit surveys.
On the one hand, hot stars\footnote{In this paper, we will follow \citet{winn:2010} and 
define hot stars as those with $\teff >6250$K. This is also roughly the temperature of the 
Kraft break for stars near the zero age main sequence (see \citet{kraft:1967} and Figure \ref{fig:teffhist}).}
have been relatively unexplored as compared to later spectral
types.  The first transiting planet was discovered by radial
velocity surveys \citep{charbonneau:2000,henry:2000}, which initially targeted only late F, G, and
iearly K stars.  Due to the magnitude range of the stars surveyed and
the choice of which candidates to follow up, the first dedicated ground-based
transit surveys \citep{alonso:2004,mccullough:2006,bakos:2007,colliercameron:2007} were also primarily sensitive to
late F, G, and early K stars.  As the value of transiting
planets orbiting later stellar types was increasingly recognized,
transit surveys began to survey lower-mass stars.  Kepler extended
the magnitude range of their target sample to fainter magnitudes \citep{gould:2003,batalha:2010}, in
order to include a significant number of M dwarfs. The Kepler K2 mission will likely survey
an even larger number of M stars than the prime mission \citep{howell:2014}. MEarth is
specifically targeting a sample of some 3,000 mid to late M dwarfs \citep{irwin:2014}.
Finally, HAT-South is surveying even fainter stars than HATNet, in
order to increase the fraction of late G, K, and even early M stars \citep{bakos:2013}.  

As a result of this focus on later spectral types, the population of
close-in, low-mass companions to hot stars is relatively poorly
assayed. This is particularly true for stars which are both hot and massive;
for example, only $6$ transiting planetary companions are
known orbiting stars with $\teff>6250$K and $M>1.5 \msun$.\footnote{According to http://exoplanets.org}
Building a larger
sample is particularly important given existing claims that the population of
planetary and substellar companions to hot and/or massive stars is
different than that of cooler and less massive stars.
In particular, there is evidence that hot Jupiters orbiting hot stars tend to
have a large range of obliquities \citep{winn:2010,schlaufman:2010, albrecht:2012}.
Based on surveys of giant stars, whose progenitors are likely to be massive
(\citet{johnson:2013}, but see \citet{lloyd:2013}), there have also been claims
that the distribution of Jovian 
planetary companions is a strong function of primary mass (\citet{bowler:2010},
but see e.g. \citet{maldonado:2013}).
Finally, there is anecdotal evidence that massive substellar
companions are more common around stars with $\teff>\sim6200$K \citep{bouchy:2011}.

The challenges posed by hot stars are primarily due to
the high rotation velocities. The large rotation velocities of hot
stars result in broad and weak lines, making precision radial
velocity difficult.  As a result, radial velocity surveys, and to a
lesser extent transit surveys, have avoided targeting, or following up
candidates from, such stars.  Furthermore, for a fixed planet radius, the
depths of planetary transits of hotter stars are shallower.  This is
exacerbated by the fact that stars with $\teff>6250$K have lifetimes that
are of order the age of the Galactic disk, and thus tend to be
significantly evolved.

However, there a number of ways in which these challenges are
mitigated for transit surveys.  First, even though the transit depths
are shallower, they are nevertheless greater than a few
millimagnitudes, and thus readily detectable for Jovian-sized
companions.  Therefore, identifying such candidate transit signals is
possible even for main-sequence stars as hot as $7000$K.  Once a candidate signal is
identified, its period can be confirmed with photometric follow-up.
With a robust ephemeris in hand, radial velocity follow-up is greatly
eased, as one is simply looking for a reflex variation with a specific
period and phase (as opposed to searching over a wide range of these
parameters, which increases the probability of false positives).  Even with
the relatively poor precision (a few $100$ \ms) of radial velocity
measurements of hot stars, it is possible to exclude stellar
companions and detect the reflex motion of relatively massive
planetary and substellar companions.

Ultimately, however, it is precisely the high rotation velocities of
hot stars that assist in robust confirmation of planetary transits, via
the Rossiter-McLaughlin effect (RM) \citep{rossiter:1924, mclaughlin:1924}.
The rotating host star allows one to measure the spectral aberration of
the absorption lines due to the small blockage of light as the planet
transits the rapidly rotating host.  The magnitude of this effect can
be directly predicted by the rotation velocity measured from the
spectrum, combined with the transit depth and shape. The RM effect can
therefore provide strong confirmation that the transit signal is due to
a planetary-sized object transiting the target star. However, for 
Jupiter-sized companions, this does not necessarily confirm the planetary
nature of the occultor, because low-mass stars, brown dwarfs, and Jovian
planets all have roughly $\sim$\rjup\ \citep{chabrier:2000}. However, 
even a crude upper limit on the Doppler amplitude of a few \kms\ can
then be used to exclude essentially all companions with masses in the 
stellar or brown dwarf regime. Thus, the Doppler upper limit, combined
with the RM measurement, essentially confirms that the companion is a planet,
i.e. that is both mass and radius are in the planetary regime.
Furthermore, the shape of the RM signal allows
one to measure the projected angle between the planet's orbital axis
and the star's rotation axis.  This projected obliquity provides clues
to the formation and evolution \citep{albrecht:2012} history of hot
Jupiters and substellar companions. This effect also provides an
independent measurement of the rotational velocity of the star.

In this paper, we describe the discovery and confirmation of a hot
Jupiter transiting the bright $V = 8.54$ star \hd, which we
designate as KELT-7b. In Section \ref{sec:discovery}, we summarize the
discovery photometric transit signal and the follow-up photometric and 
spectroscopic observations. In Section \ref{sec:analysis}, we discuss the 
analysis of the data obtained to determine stellar and planetary parameters.
Section \ref{sec:false} considers the false positive scenarios and Section  
\ref{sec:discussion} discusses the results of this analysis.

\section{Discovery and Follow-up Observations}
\label{sec:discovery}

\subsection{KELT Observations and Photometry}
\label{sec:keltphotom}

The KELT-N survey has a standard process of data reduction which will be briefly described in this section. For more information 
see \citet{siverd:2012}.
KELT-7 is in KELT-N survey field 04, which is centered on ($\alpha=05h\colon54m\colon14.71s$, $\delta=+31d\colon39m\colon55.10s;
J2000$). Field 04 was monitored from 2006 October 26 to 2011 April 1 collecting about $7800$ images. 
The KELT-7 light curve in particular had $7745$ points after a single round of iterative $3\sigma$\ outlier clipping that occurs
just after the trend filtering algorthm (TFA)\citep{kovacs:2005}. We reduced the raw survey data using a custom implementation of the 
ISIS image subtraction package \citep{alard:1998, alard:2000}, combined with point-spread-function photometry using DAOPHOT \citep{stetson:1987}.
Using proper motions from the Tycho-2 catalog \citep{hog:2000} and $J$ and $H$ magnitudes from 2MASS \citep{skrutskie:2006, cutri:2003}, we 
applied a reduced proper motion cut \citep{gould:2003} based on the implementation of methods from \citet{colliercameron:2007}.
This allowed us to select likely dwarf and subgiant stars within the field for further post-processing and analysis. We applied
the TFA to each selected light curve to remove systematic noise, followed by
a search for transit signals using the box-fitting least squares algorithm (BLS) \citep{kovacs:2002}. For both TFA and BLS
we used the versions found in the VARTOOLS package \citep{hartman:2008}.

\begin{figure}[]
\plotone{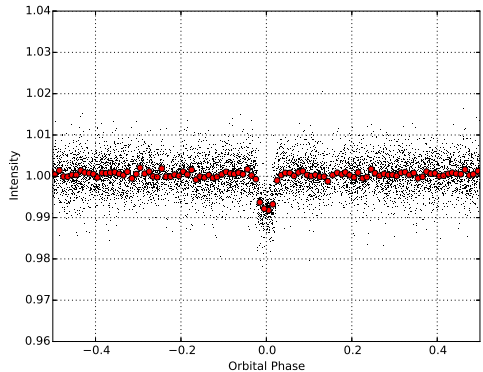}
\caption[]{Discovery light curve of KELT-7b from the KELT-N telescope. The light curve contains $7745$ observations spanning
$4.5$ years, phase folded to the orbital period of $P=\period$\ days. The red line represents the same data binned at $1$ hr
intervals in phase.
\label{fig:discoverylc}}
\end{figure}

One of the candidates from field 04 was star \hd\ / \twomass\ / \tycho, located at ($\alpha=05h\colon13m\colon10.93s$, 
$\delta=+33d\colon19m\colon05.40s; J2000$).
The star has Tycho magnitudes $\bt = \btmag$ and $\vt = \vtmag$ \citep{hog:2000} and passed our initial selection cuts.
The discovery light curve of KELT-7 is shown in Figure \ref{fig:discoverylc}. We observed a transit-like feature at a period of
$\period$ days, with a depth of about $\depthmmag$ mmag.

\subsection{Follow-up Time Series Photometry}
\label{sec:followupphotom}

\begin{figure}[]
\includegraphics[scale=0.5]{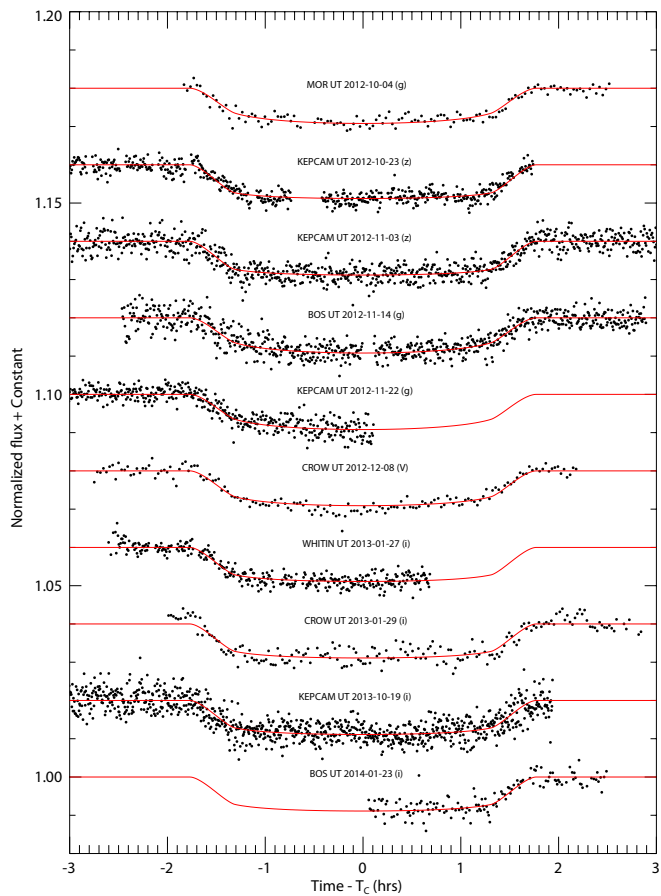}
\caption[]{
Follow-up transit photometry of KELT-7. The red over plotted line is the best fit transit model. The labels are as follows: 
MOR=University of Louisville Moore Observatory; KEPCAM=KeplerCam at the Fred Lawrence Whipple Observatory; BOS=Bryne Observatory at 
Sedgwick (LCOGT); CROW=Canela's Robotic Observatory; WHITIN=Whitin Observatory at Wellesley College
\label{fig:transitall}}
\end{figure}

We obtained follow-up time-series photometry of KELT-7 to check for false positives and better determine the transit shape.
We used the Tapir software package \citep{jensen:2013}
to predict transit events, and we obtained $10$ full 
or partial transits in multiple bands between October 2012 and January 2014. All data were calibrated and processed using the 
AstroImageJ package\footnote{http://www.astro.louisville.edu/software/astroimagej} (AIJ; K. Collins et al., in prep.) unless otherwise 
stated.

We obtained one full transit of KELT-7b in the $g$-band on UT2012-10-04 at the University of Louisville's Moore Observatory.
We used the $0.6$ m RC Optical Systems (RCOS)
telescope with an Apogee U16M 4K x 4K CCD, giving a $26^{'}$ x $26^{'}$ field of view and $0.39$ arcsec pixel$^{-1}$.
 
We used KeplerCam on the 1.2 m telescope at the Fred Lawrence Whipple Observatory (FLWO) to observe two full $z$-band transits 
on UT2012-10-23 and on UT2012-11-03. We also observed a partial $g$-band transit on UT2012-11-22. On the night of UT2013-10-19
we observed a full $i$-band transit in combination with radial velocity observations to measure the RM effect (described more
in Section \ref{sec:spectrometry}).
KeplerCam has a single 4K x 4K Fairchild CCD with 0.366 arcsec pixel$^{-1}$ and a field of view of $23.1^{'}$ x $23.1^{'}$. 
The data were reduced using procedures outlined in \citet{carter:2011}, which uses standard IDL routines.

We observed a full transit in $g$-band on UT2012-11-14 and a partial transit in $i$-band on UT2014-01-23 from the Byrne Observatory
at Sedgwick (BOS), operated by Las Cumbres Observatory Global Telescope Network (LCOGT). BOS is a 0.8 m RCOS telescope with a 3K x 2K
SBIG STL-6303E detector. It has a $14.7^{'}$ x $9.8^{'}$ field of view and $0.572$ arcsec pixel$^{-1}$. Data on the night of   
UT2012-11-14 were reduced and light curves were extracted using standard IRAF/PyRAF routines as described in \citet{fulton:2011}.
Observations from UT2014-01-23 were analyzed using custom routines written in 
GDL.\footnote{GNU Data Language; http://gnudatalanguage.sourceforge.net/}

We observed two full transits at Canela's Robotic Observatory (CROW) in Portugal. Observations were made using the 0.3 m LX200 
telescope with a SBIG ST-8XME CCD. The field of view is $28^{'}$ x $19^{'}$ and $1.11$ arcsec pixel$^{-1}$. Observations were 
taken on UT2012-12-08 and UT2013-01-29 in $V$-band and $i$-band, respectively.

We observed one partial transit in $I$-band on the night of UT2013-01-27 at the Whitin Observatory at Wellesley College. The 
observatory uses a 0.6 m Boller and Chivens telescope with a DFM focal reducer that gives an effective focal ratio of f/9.6.
The camera is an Apogee U230 2K x 2K with a 0.58 arcsec pixel$^{-1}$ scale and a $20^{'}$ x $20^{'}$ FOV. Reductions were
carried out using standard IRAF packages, with photometry done in AIJ.

\begin{figure}[]
\plotone{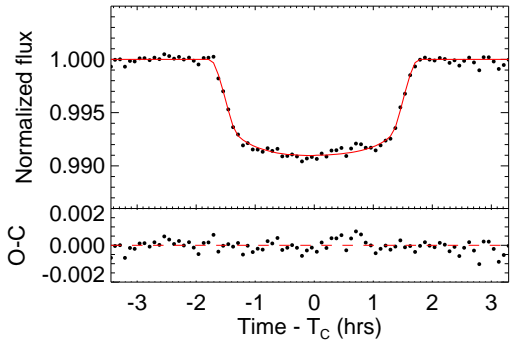}
\caption[]{
Top panel: All follow-up light curves from Figure \ref{fig:transitall}, combined and binned in 5 minute intervals. This 
light curve is not used for analysis, but is shown in order to illustrate the best combined behavior of the
light curve data set. The red curve shows the 10 transit models for each of the individual fits combined and binned in 5
minute intervals the same way as the data, with the model points connected.
Bottom panel: The residuals of the binned light curve from the binned model in the top panel.
\label{fig:transitcomb}}
\end{figure}

Figure \ref{fig:transitall} shows each transit plotted with the best fit transit model over plotted in red. Figure 
\ref{fig:transitcomb} shows the combined and binned light curve with all $10$ transits. Plots were generated during
the Global Fit analysis (See Section \ref{sec:exo}) using EXOFAST \citep{eastman:2013}.

\subsection{Spectroscopic Observations}
\label{sec:spectrometry}

We used the Tillinghast Reflector Echelle Spectrograph \citep[TRES;][]{furesz:2008}, on the 1.5\,m telescope
at the Fred Lawrence Whipple Observatory (FLWO) on Mt. Hopkins, Arizona, to obtain spectra to test false positive 
scenarios, characterize radial velocity (RV) variations and determine stellar parameters of the host star.  We obtained a total
of $64$ TRES spectra between UT 2012 January 31 and UT 2014 February 21. The exposure time varied from $90-2400$ seconds depending
on the weather conditions and the SNR we were trying to achieve. The spectra have a resolving power of
$R=44,000$ and were extracted as described by \citet{buchhave:2010}.

Initially we obtained observations near phases $0.25$ and $0.75$ in order to check for large velocity variations due to a
small stellar companion responsible for the light curve.
The spectrum appeared to be single-lined, and the velocity variation that we saw was much too small to be due to 
a stellar companion so we continued observing to get a preliminary orbit. The orbit had a fair
amount of scatter due to the rapid rotation of the host star so we stopped observing spectroscopically and
opted instead to follow-up the star photometrically to confirm the depth, shape, and period of the transit, and to search
for color-depth dependent depth variations indicative of a blended eclipsing binary.
Once we had observed multiple transits in different filters to determine that the transit depths were indeed achromatic,
we started obtaining high signal-to-noise ratio spectra to refine the orbit and to determine stellar parameters of the host star. 
Table \ref{tab:rvs} lists all of the RV data for KELT-7.


\begin{deluxetable*}{rrrrrrr}
\tablecaption{RV Observations of KELT-7\label{tab:rvs}}
\tabletypesize{\scriptsize}
\tablehead{
\colhead{Time}                         &
\colhead{Relative RV}                  &
\colhead{Relative RV error}            &
\colhead{Bisector}                     &
\colhead{Bisector error}               &
\colhead{Phase}                        & 
\colhead{SNRe\tablenotemark{a}}        \\
\colhead{$BJD_{TDB}$}                  &
\colhead{(\ms)}                        &
\colhead{(\ms)}                        &
\colhead{(\ms)}                        &
\colhead{(\ms)}                        &
\colhead{}                             &
\colhead{}                             \\
}
\startdata
 2455957.836103  &$   102 $&$   143 $&$  -114 $&$    209 $&$  -0.3053  $&$   68.8 $ \\
 2455961.748540  &$  -472 $&$   134 $&$   -37 $&$    143 $&$   1.1253  $&$   87.4 $ \\
 2455963.607625  &$   -65 $&$   144 $&$   -35 $&$    102 $&$   1.8050  $&$  122.7 $ \\
 2455964.647614  &$  -387 $&$   174 $&$   170 $&$    137 $&$   2.1853  $&$   82.2 $ \\
 2455967.702756  &$  -129 $&$    73 $&$    16 $&$    116 $&$   3.3024  $&$  101.0 $ \\
 2455968.667152  &$   -25 $&$    86 $&$   -52 $&$     61 $&$   3.6551  $&$  191.2 $ \\
 2455970.667266  &$  -135 $&$   136 $&$   -79 $&$     69 $&$   4.3864  $&$  106.3 $ \\
 2455971.657526  &$    67 $&$    95 $&$  -199 $&$     77 $&$   4.7485  $&$  161.3 $ \\
 2455982.637552  &$   -13 $&$    83 $&$   127 $&$     53 $&$   8.7633  $&$  153.3 $ \\
 2455983.610040  &$  -150 $&$    97 $&$    55 $&$    107 $&$   9.1189  $&$  174.2 $ \\
 2455984.641133  &$  -197 $&$    90 $&$    86 $&$     76 $&$   9.4959  $&$  180.2 $ \\
 2455985.610681  &$    94 $&$   104 $&$   -55 $&$    107 $&$   9.8504  $&$  122.0 $ \\
 2455986.621278  &$  -267 $&$   123 $&$  -164 $&$    156 $&$  10.2200  $&$   92.1 $ \\
 2456019.663088  &$  -273 $&$   115 $&$   108 $&$     92 $&$  22.3017  $&$   73.5 $ \\
 2456020.642102  &$  -139 $&$   107 $&$  -100 $&$     88 $&$  22.6597  $&$  125.5 $ \\
 2456027.643080  &$   -52 $&$    89 $&$  -262 $&$    119 $&$  25.2196  $&$  141.2 $ \\
 2456310.790708  &$    55 $&$   117 $&$   144 $&$     60 $&$ 128.7524  $&$  296.9 $ \\
 2456340.790829  &$     0 $&$    48 $&$    97 $&$     44 $&$ 139.7219  $&$  297.5 $ \\
 2456584.819001  &$   -45 $&$    88 $&$   102 $&$    131 $&$ 228.9507  $&$  131.4 $ \\
 2456584.826693  &$    81 $&$    64 $&$    26 $&$     71 $&$ 228.9535  $&$  127.1 $ \\
 2456584.835357  &$    62 $&$   100 $&$   -42 $&$     85 $&$ 228.9567  $&$  146.6 $ \\
 2456584.843054  &$    -1 $&$    93 $&$    75 $&$     70 $&$ 228.9595  $&$  148.1 $ \\
 2456584.852250  &$    65 $&$    85 $&$     8 $&$     95 $&$ 228.9629  $&$  144.0 $ \\
 2456584.859959  &$   -62 $&$    63 $&$    41 $&$     87 $&$ 228.9657  $&$  144.5 $ \\
 2456584.867813  &$    13 $&$    48 $&$    14 $&$     78 $&$ 228.9686  $&$  145.7 $ \\
 2456584.875811  &$     7 $&$    68 $&$     8 $&$    104 $&$ 228.9715  $&$  143.5 $ \\
 2456584.884423  &$    96 $&$   108 $&$   -39 $&$     74 $&$ 228.9746  $&$  154.8 $ \\
 2456584.892300  &$   230 $&$    96 $&$  -130 $&$     57 $&$ 228.9775  $&$  154.3 $ \\
 2456584.899864  &$   181 $&$    71 $&$  -179 $&$     70 $&$ 228.9803  $&$  155.7 $ \\
 2456584.907399  &$   265 $&$   110 $&$    33 $&$    112 $&$ 228.9830  $&$  161.2 $ \\
 2456584.915097  &$    86 $&$    82 $&$    89 $&$     61 $&$ 228.9858  $&$  155.4 $ \\
 2456584.923361  &$   146 $&$    98 $&$   138 $&$     62 $&$ 228.9889  $&$  154.1 $ \\
 2456584.930798  &$    93 $&$    81 $&$    56 $&$     43 $&$ 228.9916  $&$  153.1 $ \\
 2456584.938490  &$   -78 $&$    65 $&$    98 $&$     91 $&$ 228.9944  $&$  162.9 $ \\
 2456584.946413  &$     1 $&$    76 $&$    -9 $&$     55 $&$ 228.9973  $&$  162.7 $ \\
 2456584.956825  &$   -87 $&$   103 $&$   -80 $&$     84 $&$ 229.0011  $&$  161.8 $ \\
 2456584.964968  &$  -172 $&$    82 $&$   -77 $&$     78 $&$ 229.0041  $&$  166.5 $ \\
 2456584.972520  &$  -205 $&$   109 $&$   -70 $&$     74 $&$ 229.0068  $&$  165.2 $ \\
 2456584.980362  &$  -255 $&$    89 $&$  -143 $&$     82 $&$ 229.0097  $&$  158.9 $ \\
 2456584.987938  &$  -382 $&$    80 $&$   -23 $&$     56 $&$ 229.0125  $&$  164.0 $ \\
 2456584.996897  &$  -610 $&$   118 $&$   132 $&$     72 $&$ 229.0158  $&$  162.3 $ \\
 2456585.004641  &$  -437 $&$    81 $&$   240 $&$     53 $&$ 229.0186  $&$  165.4 $ \\
 2456585.012118  &$  -223 $&$    70 $&$   336 $&$     79 $&$ 229.0213  $&$  162.9 $ \\
 2456585.019665  &$   -86 $&$    80 $&$   207 $&$     70 $&$ 229.0241  $&$  159.9 $ \\
 2456585.027143  &$   -37 $&$   123 $&$   243 $&$     65 $&$ 229.0268  $&$  156.1 $ \\
 2456585.034609  &$  -226 $&$    78 $&$   145 $&$     70 $&$ 229.0295  $&$  156.2 $ \\
 2456638.901881  &$    26 $&$    98 $&$  -169 $&$    112 $&$ 248.7261  $&$  198.1 $ \\
 2456639.777025  &$  -123 $&$    68 $&$   -67 $&$     73 $&$ 249.0461  $&$  242.0 $ \\
 2456640.913299  &$  -115 $&$    80 $&$    90 $&$     53 $&$ 249.4616  $&$  262.4 $ \\
 2456641.702340  &$  -166 $&$    78 $&$     6 $&$     67 $&$ 249.7501  $&$  169.0 $ \\
 2456642.742449  &$  -347 $&$    66 $&$    39 $&$     92 $&$ 250.1304  $&$  258.3 $ \\
 2456693.611029  &$    72 $&$    73 $&$  -200 $&$     72 $&$ 268.7305  $&$  203.3 $ \\
 2456694.641506  &$  -208 $&$    89 $&$   -67 $&$     68 $&$ 269.1073  $&$  186.2 $ \\
 2456696.631120  &$    11 $&$    73 $&$  -149 $&$     70 $&$ 269.8348  $&$  229.6 $ \\
 2456700.710497  &$  -519 $&$   134 $&$    -2 $&$     75 $&$ 271.3264  $&$  146.0 $ \\
 2456701.756533  &$   -38 $&$    67 $&$  -206 $&$     70 $&$ 271.7089  $&$  238.9 $ \\
 2456702.691941  &$  -207 $&$    53 $&$   -53 $&$     50 $&$ 272.0509  $&$  278.9 $ \\
 2456703.670164  &$  -194 $&$    59 $&$   -35 $&$     42 $&$ 272.4086  $&$  263.7 $ \\
 2456704.671919  &$   -94 $&$    77 $&$    61 $&$     50 $&$ 272.7749  $&$  219.1 $ \\
 2456705.773764  &$  -316 $&$    76 $&$   -78 $&$     75 $&$ 273.1778  $&$  220.1 $ \\
 2456706.652092  &$   -32 $&$    68 $&$   -41 $&$     60 $&$ 273.4989  $&$  254.3 $ \\
 2456707.664631  &$     9 $&$    63 $&$   -63 $&$     37 $&$ 273.8692  $&$  238.4 $ \\
 2456708.711942  &$  -381 $&$    84 $&$    80 $&$     42 $&$ 274.2521  $&$  206.1 $ \\
 2456709.746023  &$   -21 $&$    65 $&$   -27 $&$     59 $&$ 274.6302  $&$  260.8 $ \\

\enddata
\tablenotetext{a}{
    Signal to noise per resolution element (SNRe) which takes into account the resolution of the instrument. 
    SNRe is calculated near the peak of the echelle order that includes the Mg b lines.
}
\end{deluxetable*}

Of the $64$ total spectra, $28$ were taken on the night of UT 2013 October 19 to measure the RM effect and determine the projected
obliquity of the system. Simultaneous data were taken using the
TRES spectrograph on the $1.5$ m telescope and photometric data using KeplerCam on the $1.2$ m telescope both atop Mt. 
Hopkins in AZ. We collected $28$ RV spectra with a $9$ minute exposure cadence and signal-to-noise ratio ranging from $127$ to $165$ 
per resolution element. For the spectroscopic observations we began collecting data an hour and a half before ingress but we only obtained $2$ 
observations after egress due to morning twilight. Photometric observations were gathered starting two hours prior to ingress
until morning twilight which occurred about 10 minutes after egress. We obtained a total of $997$ KeplerCam observations at 
an exposure time of $2$ seconds and a slight defocus of the image because of the brightness of the star. 

\subsection{Adaptive Optics Observations}
\label{sec:ao}

We obtained adaptive optics (AO) imagery for KELT-7 on UT 2014 August 17 using the NIRC2 (instrument PI: Keith Matthews) with the Keck II Natural Guide Star (NGS) AO system~\citep{wizinowich:2000}. We used the narrow camera setting with a plate scale of 10 mas pixel$^{-1}$. The setting provides a fine spatial sampling of the instrument point spread function (PSF). The observing conditions were good, with seeing of $0.5^{\prime\prime}$. KELT-7 was observed at an airmass of 1.31. We used a Br-$\gamma$ filter to acquire images with a 3-point dither method. At each dither position, we took an exposure of 0.5 second per coadd and 20 coadds. The total on-source integration time was 30 sec.

\begin{figure}
\begin{center}
\includegraphics[angle=0, width=0.5\textwidth]{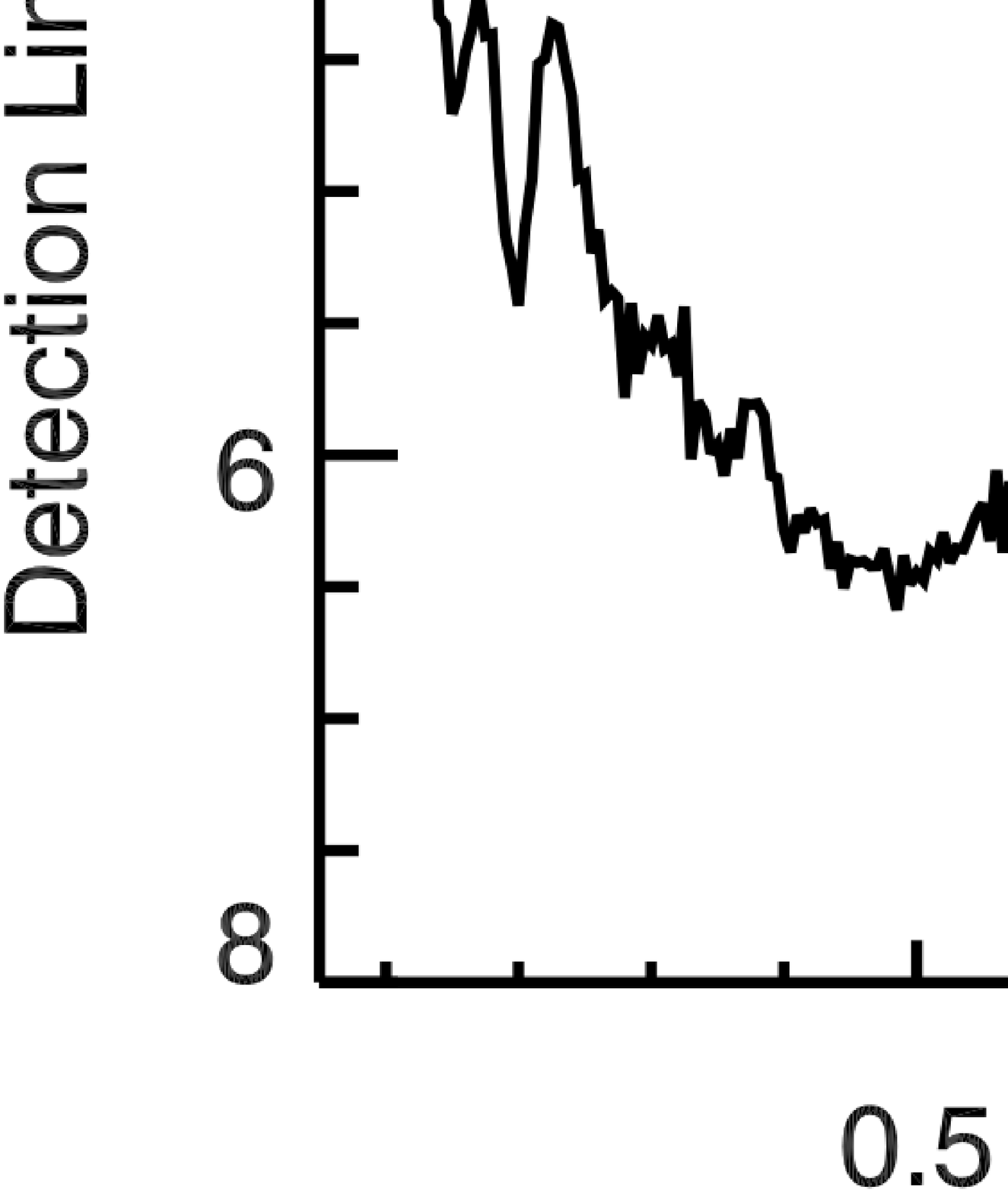} 
\caption{Top: Br-$\gamma$ AO image for KELT-7 (HD 33643) in Br-$\gamma$ ($\lambda=2.1654$ mm). North is up, east is left. The horizontal bar is 1$^{\prime\prime}$. No nearby companions or background sources were detected. Bottom: $5\sigma$ detection limit as a function of angular separation. Detection limits at $0.1^{\prime\prime}$, $0.2^{\prime\prime}$, $0.5^{\prime\prime}$, and $1.0^{\prime\prime}$ are also given in Section \ref{sec:ao}.
\label{fig:ao}}
\end{center}
\end{figure}

The raw NIRC2 data were processed using standard techniques to replace bad pixels, flat-field, subtract thermal background, align and co-add frames. We did not find any nearby companions or background sources at the 5-$\sigma$ level (Fig. \ref{fig:ao}). We calculated the 5-$\sigma$ detection limit as follows. We defined a series of concentric annuli centered on the star. For the concentric annuli, we calculated the median and the standard deviation of flux for pixels within these annuli. We used the value of five times the standard deviation above the median as the 5-$\sigma$ detection limit. The 5-$\sigma$ detection limits are $\Delta$mag=2.5 mag, 5.4 mag, 6.4 mag, and 7.3 mag for $0.1^{\prime\prime}$, $0.2^{\prime\prime}$, $0.5^{\prime\prime}$, and $1.0^{\prime\prime}$, respectively.

\section{Analysis}
\label{sec:analysis}

\subsection{Stellar Parameters}
\label{sec:spc}


\begin{deluxetable*}{lcccr}
\tablecaption{Stellar Properties of KELT-7\label{tab:stellar}}
\tabletypesize{\scriptsize}
\tablehead{
\colhead{Parameter} &
\colhead{Description} &
\colhead{Value} &
\colhead{Source} &
\colhead{Reference\tablenotemark{a}} \\
}
\startdata
Names & & \bd & &  \\
      & & \tycho & &  \\
      & & \twomass & &  \\
      & & \gsc & &  \\
      & & \hd & &  \\
$\alpha_{J2000}$ & &$ 05\ 13\ 10.93$ & Tycho-2 & $1$ \\
$\delta_{J2000}$ & &$+33\ 19\ 05.40$ & Tycho-2 & $1$ \\
$NUV_{GALEX}$ & &$13.330 \pm 0.905$ & GALEX & $2$ \\ 
$B_{T}$ & &$9.074 \pm 0.030$ & Tycho-2 & $1$ \\  
$V_{T}$ & &$8.612 \pm 0.030$ & Tycho-2 & $1$ \\  
$V$ & &$8.540 \pm 0.030$ & SKY2000 & $3$ \\
$B$ & &$8.970 \pm 0.030$ & SKY2000 & $3$ \\
$U$ & &$9.010 \pm 0.030$ & SKY2000 & $3$ \\
$I_C$ & &$8.129 \pm 0.051$ & TASS & $4$ \\
$J$ & & $7.739 \pm 0.030$ & 2MASS & $5$ \\
$H$ & & $7.580 \pm 0.042$ & 2MASS & $5$ \\
$K$ & & $7.543 \pm 0.030$ & 2MASS & $5$ \\
$WISE1$& &$10.179 \pm 0.050 $ & WISE & $6$ \\
$WISE2$& &$10.844 \pm 0.050 $ & WISE & $6$ \\
$WISE3$& &$12.766 \pm 0.180 $ & WISE & $6$ \\
$WISE4$& &$13.741 \pm 0.123 $ & WISE & $6$ \\
$\mu_{\alpha}$ & Proper Motion in RA (mas yr\textsuperscript{-1}) & $10.40 \pm 0.70$ & UCAC4 & $7$ \\
$\mu_{\delta}$ & Proper Motion in DEC (mas yr\textsuperscript{-1}) & $-49.70 \pm 0.60$ & UCAC4 & $7$ \\
$U\tablenotemark{b}$ & $\kms$ & $-33.5 \pm 0.2$ & This paper & \\
$V$ & $\kms$ & $-9.7 \pm 1.8$ & This paper & \\
$W$ & $\kms$ & $-8.4 \pm 0.9$ & This paper & \\
$d$ & Distance (pc) & $ 129 \pm 8 $ & This paper & \\
& Age (Gyr) & $ 1.3 \pm 0.2 $ & This paper & \\
$A_{V}$ & Visual extinction & $ 0.13\pm 0.04 $ & This paper & \\

\enddata
\tablenotetext{a}{References: (1)\citet{hog:2000}; (2)\citet{martin:2005}; (3)\citet{myers:2001}; (4)\citet{richmond:2000};
(5)\citet{cutri:2003, skrutskie:2006}; (6)\citet{wright:2010, cutri:2012}; (7)\citet{zacharias:2013}}
\tablenotetext{b}{Positive U is in the direction of the Galactic center.}
\end{deluxetable*}

Using the Spectral Parameter Classification (SPC) \citep{buchhave:2012} technique,
with \teff, \logg, \meh, and \vsini\ as free parameters, we obtained stellar parameters of KELT-7
from the $64$ TRES spectra. SPC cross correlates an observed spectrum against a grid of synthetic spectra based
on Kurucz atmospheric models \citep{kurucz:1992}. 
The weighted average results are: \teff $= 6779 \pm  50$ K, \logg $= 4.23 \pm 0.10$, 
\meh $= 0.12 \pm 0.08$, and \vsini $= 73.2 \pm 0.5$ \kms. \meh\ was substituted for \feh\ in this analysis but 
we do not believe that this will affect the results. The weighted mean values were calculated by taking an 
average of the stellar parameters that were calculated for each spectra individually, and weights them according
to the cross-correlation function (CCF) peak height.
\footnote{SPC compares the observed spectra against a library of
synthetic spectra calculated with the same mix of metals as the Sun. Since it uses all the lines in the observed
spectra in the wavelength region covered by the library, the metallicity \meh\ is the same as \feh\ only if the mix 
of metals in the target star is the same as the Sun.}

\subsection{Radial Velocity Analysis}
\label{sec:specanalysis}

\begin{figure}
\plotone{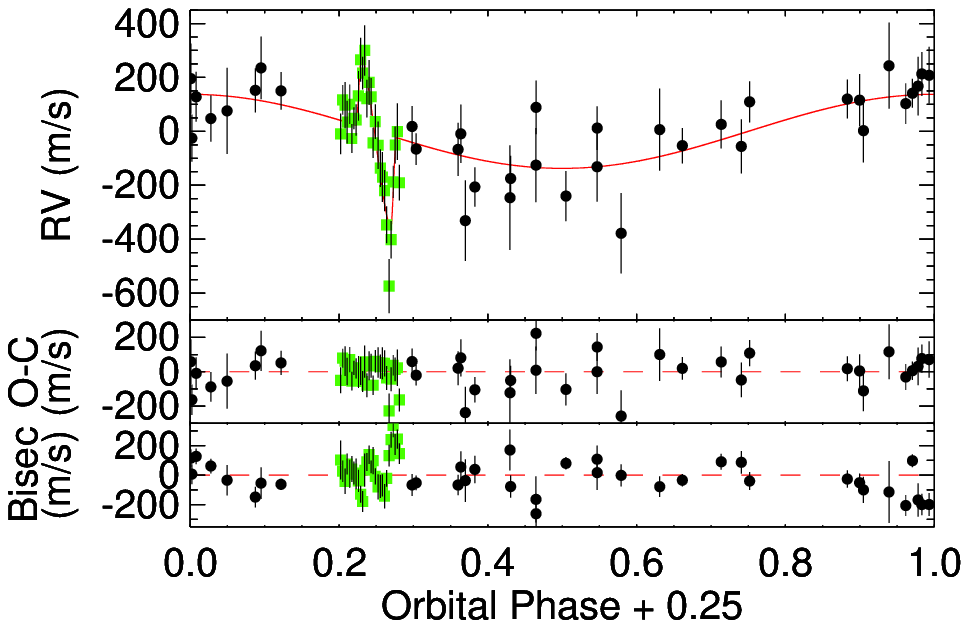}
\caption{
TRES radial velocities of KELT-7. The green squares represent data taken on the night of the RM event, while the black
circles are data that were not taken during transit. The phases have been shifted so that a phase of $0.25$ corresponds to
the time of the primary transit, $T_{C}$. Top panel: RV observations phased to the best orbital model with eccentricity fixed to
zero and with no linear trend, shown in red. The predicted RM effect in the model shown incorporates the best fit model
where $\lambda = 9.7 \pm 5.2$ degrees. Middle panel: Residuals
of the RV observations to our circular orbital fit. Bottom panel: Bisector span of the RV observations as a function of 
phase. 
}
\label{fig:rvs}
\end{figure}

The relative RVs were derived by cross-correlating the spectra against the strongest observed spectrum from the wavelength
range $4250$ - $5650 \AA$.
Figure \ref{fig:rvs} shows the best-fit orbit and computed bisectors with residuals. The best-fit orbit is a result of
the EXOFAST Global Fit (See Section \ref{sec:exo}) assuming a fixed eccentricity of zero. 
The bisector analysis of the RVs taken out of transit showed no indication that the bisector spans were in phase with the photometric
ephemeris but the RMS was large due to the high \vsini. Despite the higher \vsini\ and resulting poorer
precision, we were ultimately able to detect the reflex signal at high confidence (roughly $7\sigma$). We do see a correlation between the 
bisectors and RVs taken during the transit due to the RM effect. The relative RVs and bisector values are listed in Table \ref{tab:rvs}.

\subsection{Rossiter-McLaughlin Analysis}
\label{sec:rm}

We performed an analysis of the RM data separately from the global fit analysis (see Section \ref{sec:exo}). To model the RM effect, 
we used parameter estimation and 
model fitting protocols as described in \citet{sanchis:2013} and \citet{albrecht:2012}. The code implements formulas from \citet{hirano:2011}, 
using the loss of light calculated from transit parameters and planet position as inputs. The transit data from the night of the RM event were
used to determine the time of transit and transit parameters $b$, $\rsa$, $\rplrs$. Additional free parameters are $\vsini$ and 
$\lambda$ to describe the amplitude and shape of the signal, and a slope $\dot{\gamma}_{\rm RM}$ and offset $\gamma_{\rm RM}$ to describe the orbital motion 
of the star. $\lambda$ is the angle on the sky measured clockwise from the sky-projection of the orbit angular momentum vector, to the 
sky-projection of the stellar angular momentum vector \citep{ohta:2005}.
The uncertainty of the model parameters were estimated using a Markov Chain Monte Carlo (MCMC) algorithm, where the number of chains 
was large enough to guarantee the robustness of the final values. 

\begin{figure}[]
\plotone{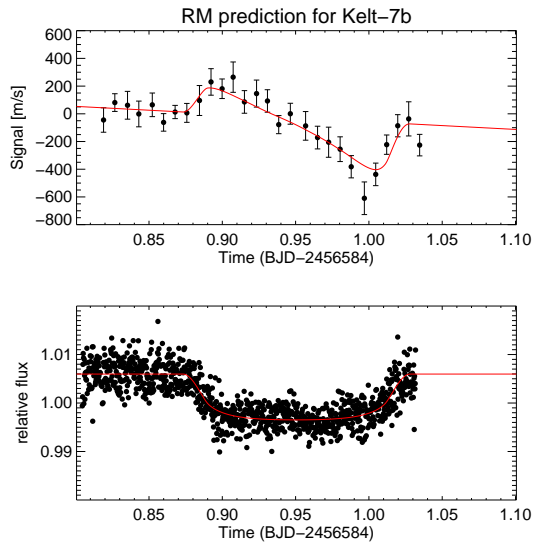}
\caption[]{RM results from our independent analysis. Spectroscopic and photometric data are from UT2013 October 19. 
Top Panel: RV observations with the best RM fit shown in red. Bottom Panel:
Photometric transit data with the best fit shown in red. Data are plotted in time for comparison.
\label{fig:rmind}
}
\end{figure}

The results from the analysis show that the sky-projected obliquity is $\lambda=4.1_{-7.7}^{+7.9}$ degrees. The analysis also gives us an 
independent measure of the projected rotational velocity, $\vsini=66_{-19}^{+21}$\ $\kms$, which is consistent with the SPC analysis (see Section 
\ref{sec:spc}). The $\gamma_{\rm RM}$ offset was determined to be $-54_{-33}^{+34}$\ $\ms$. 
We can also use the result from the out of transit acceleration $\dot{\gamma}_{\rm RM}=-671_{-340}^{+346}$\ $\msday$\ to 
estimate the velocity semi-amplitude due to the planet. Using the orbital period and assuming a circular orbit, we calculate
$K_{RV}=292\pm146$\ $\ms$.  The results from this analysis are shown in Figure \ref{fig:rmind}.

\subsection{Time-series Spectral Line Profile}
\label{sec:lineprofile}

\begin{figure}[]
\includegraphics[angle=0, width=0.5\textwidth]{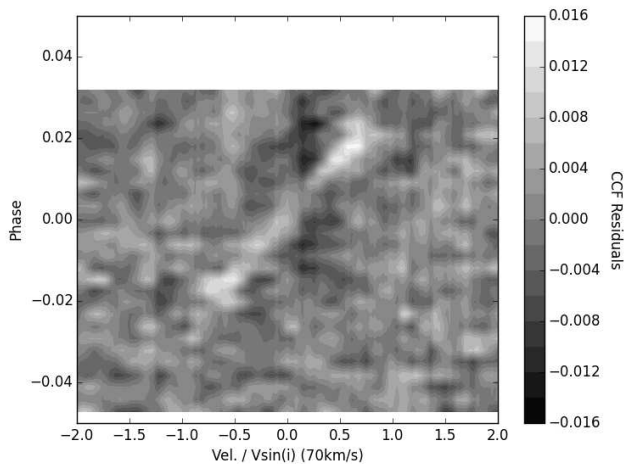}
\caption[]{Time series of the residual average spectral line profile for data taken on UT2013 October 19. The bright
white feature is caused by the planet transiting the star.
}
\label{fig:river}
\end{figure}

The overall starlight that is blocked by the planet during transit will appear as a bump in the rotational broadening function 
\citep{colliercameron:2010}. Each spectrum taken on the night of the RM event was
cross-correlated against a non-rotating template. The CCFs of the out-of-transit data were median combined to create a
master OOT CCF. The master was then subtracted from all of the in- and out-of-transit CCFs. Figure \ref{fig:river} is a greyscale
plot showing the results.  The bright white feature increasing in velocity with time is caused by the planet as it transits the star. This 
feature is what we would expect to see for a system that has low orbital obliquity: the planet crosses from the blue-shifted side of the stellar
disk to the red-shifted side, and the center of the transit occurs at a \vsini near zero.

\subsection{EXOFAST Global Fit}
\label{sec:exo}

We used a custom version of EXOFAST \citep{eastman:2013} to determine a global fit of the system. EXOFAST does a simultaneous
MCMC analysis of the photometric and spectroscopic data, including constraints on the stellar parameters  
of \mstar\ and \rstar\ from the empirical Torres relations \citep{torres:2010} or Yonsei-Yale (YY) evolutionary  
models \citep{demarque:2004}, to derive system parameters. This method is similar to that described in detail in \citet{siverd:2012},
but we note a few differences below \footnote{In the EXOFAST analysis, which includes the modeling of filter-specific limb darkening
parameters of the transit, we employ the transmission curves defined for the primed SDSS filters rather than the unprimed versions. 
We expect any differences due to that discrepency to be well below the precision of all our observations in this paper and of the
limb darkening tables from \citet{claret:2011}.}.

\begin{figure}[]
 \plotone{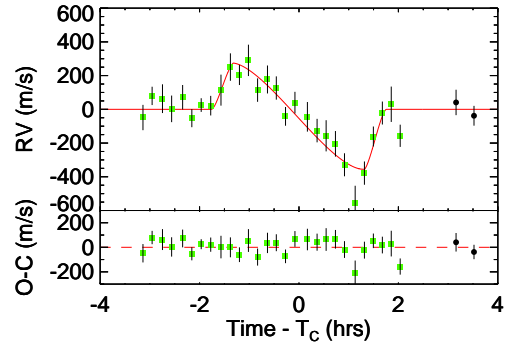}
 \caption[]{RM results from the global EXOFAST fit with the reflex velocity subtracted out. 
Top panel: RV data (green points) from UT 2013 October 19 with the best fit model shown in red.
The two black points were taken on different nights and not included in the fit. The shape of the RM signal implies that the 
projected obliquity of the host star with respect to the planet is small.  Bottom panel: The residuals of the data to the RM fit.
 \label{fig:rm}
 }
 \end{figure}

As initial inputs for EXOFAST we included as priors the orbital period $P=\period \pm 0.000004$ days from the KELT-N data and 
the host star effective temperature $\teff=\teffspc\pm50$ K, metallicity $\feh=\fehspc\pm0.08$, and stellar surface gravity $\logg=
\loggspc\pm0.10$ from TRES spectroscopy. The priors were implemented as a $\chi^{2}$ penalty in EXOFAST (see \citet{eastman:2013} 
for details). In fitting the TRES RVs independently to a Keplerian model we did not detect a significant slope in the RVs (i.e., due to 
an additional long-period companion), and we therefore did not include this as a free parameter in our final fits.

We used the AIJ package to determine detrending parameters for the light curves, such as corrections for airmass and meridan flip. 
AIJ allows interactive detrending capabilities. Once
we determined the detrending parameters for each light curve, we fit the transit light curves using EXOFAST. All light curves are
detrended by airmass while the CROW light curve from UT2013 January 29 was also detrended by meridan flip and average FWHM in the image. 
The raw data with detrending parameters were used as the input for EXOFAST, and final detrending was done in EXOFAST.

There were a few other considerations when running the global fit. First, we had to choose whether to include just the full transits,
with an ingress and an egress, or to include all transits including the partial transits that were missing an ingress or an 
egress. Second, we had the option of allowing the orbital eccentricity and argument of periastron to float free or to fix them to zero 
and force a circular orbit. Third, we had to choose between constraining the mass-radius relationship using the Torres relations or 
by using the Yonsei-Yale stellar models. Finally, we had the option to include the RM observations and fit the RM RVs as part of 
the global fit.

For the initial runs, we chose to use only the full transits and 
the non-RM RVs to ensure convergence. We also chose to fix the eccentricity to zero and set the constraint on the mass-radius relationship
using the Torres relations. Once the fit converged, we ran it again but changed the constraint on the mass-radius relationship to
the YY stellar models. We found the final parameters were in agreement within the uncertainties. This gave us confidence to start 
adding the partial transits. We initially had trouble getting the fit to converge with all the light curves and found that we needed 
to add prior width constraint on the transit timing variation (TTV) and the baseline flux for the partial transits.  
We then released the prior on eccentricity and let it float free. The result was $e=0.013_{-0.010}^{+0.022}$ which was consistent with 
a circular orbit.

\begin{deluxetable}{lccccl}
\tablewidth{0pt}
\tabletypesize{\footnotesize}
\tablecaption{Transit Times for KELT-7}
\tablehead{\colhead{Epoch} & \colhead{$T_{C}$} & \colhead{Error} & \colhead{O-C} & \colhead{O-C/Error} & \colhead {Observatory}\\ 
\colhead{} & \colhead{($BJD_{TDB}$)} & \colhead{sec} & \colhead{sec} & \colhead{} & \colhead{}}
\startdata
 -55 &$ 2456204.817057 $&$ 64 $&$   5.65 $&$  0.09 $&$ MOR $ \\
 -48 &$ 2456223.959470 $&$ 31 $&$ -83.91 $&$ -2.70 $&$ KEPCAM $ \\
 -44 &$ 2456234.898861 $&$ 42 $&$ -59.97 $&$ -1.42 $&$ KEPCAM $ \\
 -40 &$ 2456245.839584 $&$ 50 $&$  79.05 $&$  1.56 $&$ BOS $ \\
 -37 &$ 2456254.045118 $&$ 63 $&$ 182.60 $&$  2.88 $&$ KEPCAM $ \\
 -31 &$ 2456270.451621 $&$ 55 $&$  -4.72 $&$ -0.08 $&$ CROW $ \\
 -13 &$ 2456319.678871 $&$ 59 $&$ 102.16 $&$  1.72 $&$ WHITIN $ \\
 -12 &$ 2456322.413721 $&$ 56 $&$ 108.34 $&$  1.92 $&$ CROW $ \\
  84 &$ 2456584.950978 $&$ 47 $&$ -19.45 $&$ -0.41 $&$ KEPCAM $ \\
 119 &$ 2456680.667558 $&$ 87 $&$ -77.12 $&$ -0.88 $&$ BOS $ \\

\enddata
\label{tab:ttvtimes}
\end{deluxetable}

The last step of the process was to include the RM velocities in the combined global fit. The RM data were modeled using the \citet{ohta:2005}
analytic approximation.  At each step in the Markov Chain, we interpolated the linear limb darkening tables of 
\citet{claret:2011} based on the chain's value for $\logg$, $\teff$, and $\feh$\ to derive the linear limb darkening coefficient, $u$. 
Our ground-based observations generally do not constrain the two quadratic coefficients well enough to yield unique fits to both parameters. 
The uncertainties associated with this value are not true uncertainties because they include the covariances with the other parameters. For this
reason we chose not to include the linear limb darkening coefficient, $u$, in Table \ref{tab:median}.
The RM RVs were allowed a velocity offset ($\gamma_{RM}$) separate from the non-RM RV dataset velocity zeropoint ($\gamma_{RV}$). We allowed for this 
because stars have intrinsic jitter \citep{albrecht:2012, winn:2006} that can be significant in rapidly rotating stars F-stars. 
\citet{cegla:2014} have suggested that F-stars have been found to have more vigorous convective motions despite being magnetically inactive,
and that the RV jitter is strongly correlated with the granulation flicker. We find that the offset between our $\gamma_{RM}$ and $\gamma_{RV}$ 
values of $\sim100\ms$ is comparable to the RMS residual of the non-RM RV data. We also find the $\gamma_{\rm RM}$ offset determined in the 
global fit ($\gamma_{\rm RM} = -35 \pm -19$\ $\ms$) to be consistent with the $\gamma_{\rm RM}$ offset determined in Section \ref{sec:rm}.
The results from the EXOFAST RM fit are shown in Figure \ref{fig:rm}. 

For our final fits, we included all full and partial transits, 
all the radial velocities including the RM velocities, and we assumed a circular orbit with no RV slope.
The stellar and planetary values derived using the YY stellar models and using the Torres relation are shown in Table \ref{tab:median} 
for comparison. We chose to adopt the YY model as our fiducial values. We find that the spin-orbit alignment
$\lambda=9.7\pm5.2$\ degrees, and our velocity semi-amplitude, $K_{RV}=138\pm19$\ $\ms$, determined from the global fit both
agree with our independent solution discussed in Section \ref{sec:rm}. We also find that the resulting
$\vsini$ from the global fit ($\vsini = 65_{-5}^{+6}$\ $ \kms$) is in close agreement with the results from SPC and our independent RM analysis.

\begin{figure}[]
\plotone{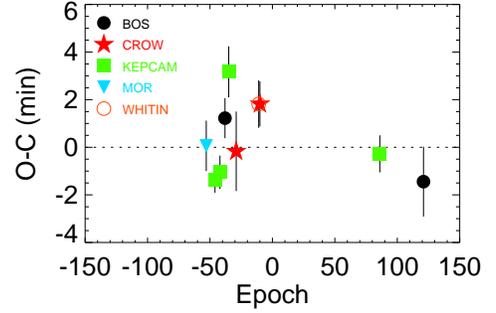}
\caption[]{The residuals of the transit times from the best-fit ephemeris. The transit times are given in Table \ref{tab:ttvtimes}. 
The WHITIN observations at epoch -13 are hidden behind the CROW observations at epoch -12.
\label{fig:ttv}
}
\end{figure}

The TTVs for all follow-up transits are shown in Figure \ref{fig:ttv}.
The global fit $T_{C}$ and $P$ were constrained only by the RV data and the priors imposed from the KELT discovery data. Using 
the follow-up transit light curves to constrain the ephemeris in the global fit would artificially reduce any observed TTV
signal.  As part of the global analysis, we fit as a free parameter a transit center time $T_{C}$ for each transit shown in 
Table \ref{tab:ttvtimes}.
A straight line was fit to all mid-transit times in Table \ref{tab:ttvtimes}, and shown in 
Figure \ref{fig:ttv}, to derive a separate ephemeris from only the transit data. We find $T_{\circ}=2456355.229809\pm0.000198$,
$P=2.7347785\pm0.0000038$, with a $\chi^{2}$\ of $27.58$ and $8$ degrees of freedom. While the $\chi^{2}$\ is much larger than
one might initally expect, this is likely due to systematics in the transit data from the ground-based photometry. Properly
removing systematics in the partial transit data would be difficult, so we are therefore not convinced that this is evidence for
TTVs. We were careful to check that all timestamps were in $BJD_{TDB}$ time system using \citet{eastman:2010} to convert timestamps. 
Further studies would be required to rule out TTVs.

\begin{deluxetable*}{lccc}
\tablewidth{0pt}
\tablecaption{Median values and 68\% confidence interval determined from EXOFAST global fit for KELT-7b}
\tablehead{\colhead{Parameter} & \colhead{Units} & \colhead{YY-isocrone Values (adopted)} & \colhead{Torres Values}}
\startdata
\sidehead{Stellar Parameters:}
$M_{*}$ &Mass (\msun) & $1.535_{-0.054}^{+0.066}$&  $1.483_{-0.068}^{+0.069}$\\
$R_{*}$ &Radius (\rsun) & $1.732_{-0.045}^{+0.043}$&  $1.715\pm0.049$\\
$L_{*}$ &Luminosity (\lsun) & $5.73_{-0.36}^{+0.37}$&     $5.61_{-0.37}^{+0.39}$\\
$\rho_*$ &Density (cgs) & $0.419_{-0.025}^{+0.027}$&  $0.415_{-0.026}^{+0.029}$\\
$\log{g_*}$ &Surface gravity (cgs) & $4.149\pm0.019$&            $4.140\pm0.019$\\
$\teff$ &Effective temperature (K) & $6789_{-49}^{+50}$&         $6789\pm49$\\
$\feh$ &Metallicity & $0.139_{-0.081}^{+0.075}$&  $0.113_{-0.083}^{+0.080}$\\
$v\sin{I_*}$\tablenotemark{a} &Rotational velocity ($km\,s^{-1}$)& $65.0_{-5.9}^{+6.0}$&    $65.4_{-5.8}^{+5.9}$\\
$\lambda$ &Spin-orbit alignment (degrees) & $9.7\pm5.2$&                $9.5_{-5.1}^{+5.2}$\\
\sidehead{Planetary Parameters:}
$P$ &Period (days) & $2.7347749\pm0.0000039$&         $2.7347750_{-0.0000039}^{+0.0000040}$\\
$a$ &Semi-major axis (AU) & $0.04415_{-0.00052}^{+0.00062}$& $0.04364_{-0.00068}^{+0.00067}$\\
$M_{P}$ &Mass (\mj) & $1.28\pm0.18$&                   $1.25\pm0.18$\\
$R_{P}$ &Radius (\rj) & $1.533_{-0.047}^{+0.046}$&        $1.514_{-0.050}^{+0.051}$\\
$\rho_{P}$ &Density (cgs) & $0.442_{-0.068}^{+0.073}$&        $0.446_{-0.069}^{+0.074}$\\
$\log{g_{P}}$ &Surface gravity & $3.131_{-0.068}^{+0.061}$&        $3.130_{-0.068}^{+0.060}$\\
$T_{eq}$ &Equilibrium temperature (K) & $2048\pm27$&                     $2051_{-27}^{+28}$\\
$\Theta$ &Safronov number & $0.0480_{-0.0067}^{+0.0069}$&    $0.0486\pm0.0068$\\
$\fave$ &Incident flux (\fluxcgs) & $4.00_{-0.20}^{+0.21}$&          $4.02_{-0.21}^{+0.22}$\\
\sidehead{RV Parameters:}
$T_C$ &Time of inferior conjunction (\bjdtdb) & $2456223.9592\pm0.0017$                    & $2456223.9591\pm0.0017$\\
$K_{RV}$ &RV semi-amplitude ($m\,s^{-1}$) & $138\pm19$                                 & $138\pm19$\\
$K_{RM}$ &RM semi-amplitude ($m\,s^{-1}$) & $542_{-50}^{+51}$                          & $543_{-49}^{+51}$\\
$M_P\sin{i}$ &Minimum mass (\mj) & $1.28\pm0.18$                              & $1.24\pm0.18$\\
$M_{P}/M_{*}$ &Mass ratio & $0.00080\pm0.00011$                        & $0.00081\pm0.00011$\\
$\gamma_{RV}$ &RV velocity zeropoint $m\,s^{-1}$ & $-133\pm15$                                & $-133\pm15$\\
$\gamma_{RM}$ &RM velocity zeropoint $m\,s^{-1}$& $-35\pm19$                                 & $-34\pm19$\\
$f(m1,m2)$ &Mass function (\mj) & $0.00000080_{-0.00000028}^{+0.00000038}$   & $0.00000080_{-0.00000029}^{+0.00000037}$\\
\sidehead{Primary Transit Parameters:}
$R_{P}/R_{*}$ &Radius of the planet in stellar radii & $0.09097_{-0.00064}^{+0.00065}$          & $0.09074_{-0.00066}^{+0.00067}$\\
$a/R_*$ &Semi-major axis in stellar radii & $5.49_{-0.11}^{+0.12}$                   & $5.47\pm0.12$\\
$i$ &Inclination (degrees) & $83.76_{-0.37}^{+0.38}$                  & $83.72_{-0.39}^{+0.40}$\\
$b$ &Impact parameter & $0.597_{-0.025}^{+0.022}$                & $0.599_{-0.026}^{+0.023}$\\
$\delta$ &Transit depth & $0.00828\pm0.00012$                      & $0.00823\pm0.00012$\\
$T_{\circ}$ &Best-fit linear ephemeris from transits (\bjdtdb) & $2456355.229809\pm0.000198$                    & $2456352.495016\pm0.000191$\\
$P_{Transit}$ &Best-fit linear ephemeris period from transits (days) & $2.7347785\pm0.0000038$&         $2.7347795\pm0.0000037$\\
$T_{FWHM}$ &FWHM duration (days) & $0.12795\pm0.00046$                      & $0.12821\pm0.00047$\\
$\tau$ &Ingress/egress duration (days) & $0.01835_{-0.00089}^{+0.00092}$          & $0.01840_{-0.00093}^{+0.00096}$\\
$T_{14}$ &Total duration (days) & $0.14630_{-0.00092}^{+0.00097}$          & $0.14662\pm0.00098$\\
$P_{T}$ &A priori non-grazing transit probability & $0.1655\pm0.0034$                        & $0.1662_{-0.0036}^{+0.0035}$\\
$P_{T,G}$ &A priori transit probability & $0.1987_{-0.0042}^{+0.0043}$             & $0.1993_{-0.0045}^{+0.0044}$\\
\sidehead{Secondary Eclipse Parameters:}
$T_{S}$ &Time of eclipse (\bjdtdb) & $2456222.5918\pm0.0017$                  & $2456222.5918\pm0.0017$

\enddata
\label{tab:median}
\tablenotetext{a}{We adopted the SPC value for $\vsini$ ($73$ \kms) as our fiducial value since the EXOFAST RM analysis is not 
designed to model rapidly rotating stars.}
\end{deluxetable*}

\subsection{Evolutionary Analysis}
\label{sec:evo}

\begin{figure}[]
\plotone{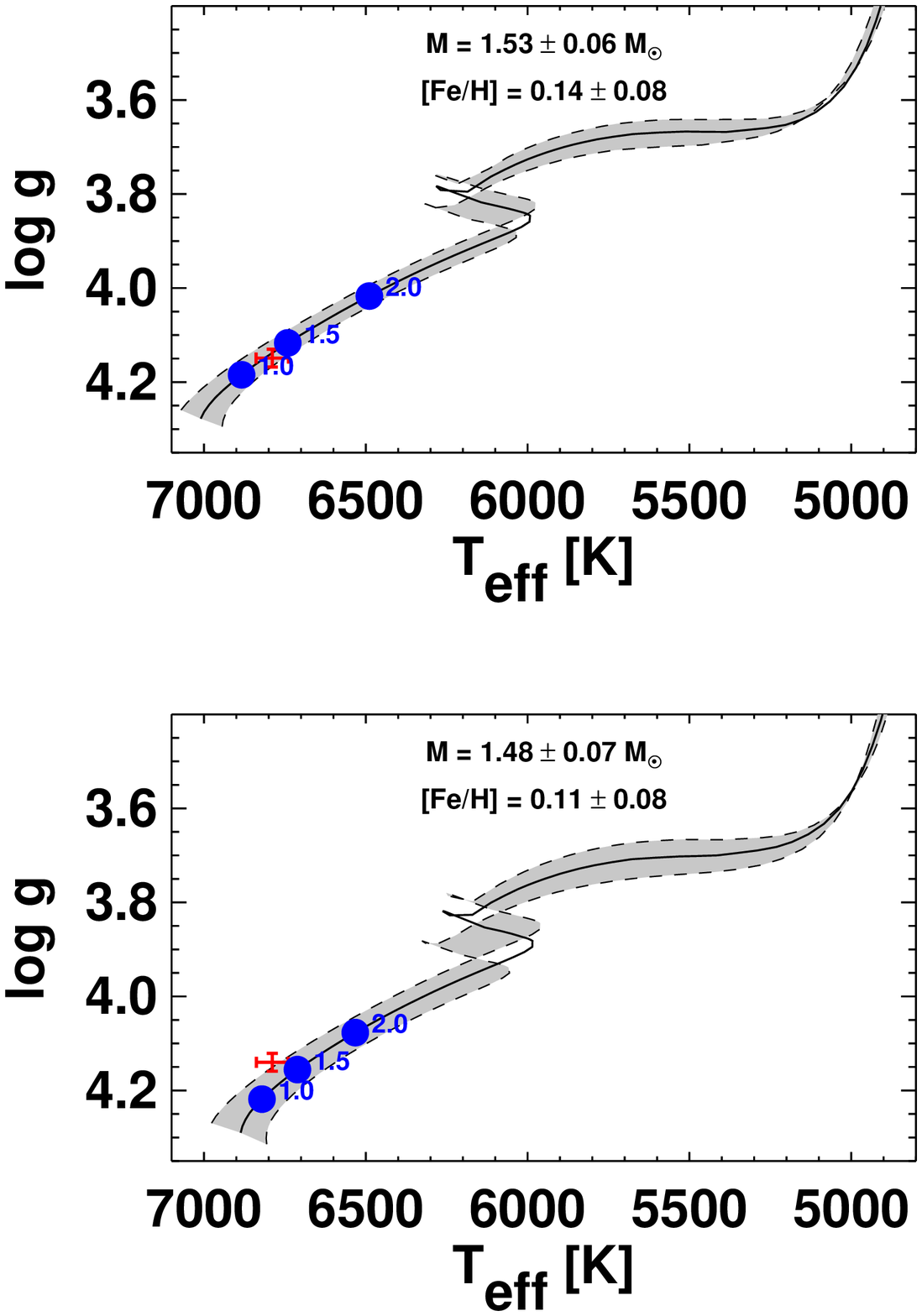}
\caption[]{Theoretical HR diagrams based on Yonsei-Yale stellar evolution models \citep{demarque:2004}. The solid
lines represent the evolutionary tracks for the best-fit values of the mass and metallicity of the host star from
the global fits using the Yonsei-Yale constraints (top panel) and the Torres constraints (bottom panel) as described 
in Section \ref{sec:exo}. The tracks for the extreme range of $1\sigma$ uncertainties on $\mstar$ and $\feh$  are shown
as dashed lines, backeting the $1\sigma$ range shown in grey. The red crosses show $\teff$ and $\logg$ from the EXOFAST global fit analysis.
the blue dots represent the location of the star for various ages in Gyr. We adopt the Yonsei-Yale constrained global
fit represented in the top panel resulting in an estimated age of $1.3 \pm 0.2$ Gyr, where we note
the uncertainty does not include possible systematic errors in the adopted evolutionary tracks.
\label{fig:iso}
}
\end{figure}

We use $\teff$, $\logg$, stellar mass, and metallicity derived from the EXOFAST global fits (see Section \ref{sec:exo} 
and Table \ref{tab:median}), in combination with  the theoretical evolutionary tracks of the Yonsei-Yale stellar models
\citep{demarque:2004}, to estimate the age of the KELT-7 system. We have not directly applied a prior on the age, but 
rather have assumed uniform priors on $\feh$, $\logg$, and $\teff$, which translates into non-uniform priors on the age.
Figure \ref{fig:iso} shows the theoretical HR diagram ($\logg$ vs. $\teff$) with evolutionary tracks for masses 
corresponding to the $\pm1\sigma$ extrema in estimated uncertainty. We adopt the Yonsei-Yale constrained global fit 
represented in the top panel. The estimated stellar mass (and secondarily the
metallicity) define the model stellar evolutionary track from which the age is inferred in the HR diagram.
Within the $1\sigma$ uncertainties on the observed $\teff$, $\logg$, and $\feh$, the 
Yonsei-Yale evolutionary track gives an inferred age of $1.3 \pm 0.2$ Gyr.
The bottom panel of Figure \ref{fig:iso} is shown as a comparison using the Torres constrained 
global fit values \citep{torres:2010}. The Torres model provides empirical relationships between observed stellar parameters  \teff, 
\logg, and \feh, and stellar mass and radius (see Table \ref{tab:median}). From this comparison we see that the age estimate 
quoted from the YY stellar model estimate is consistent with that inferred from the Torres model estimated parameters to
within $1 \sigma$.

\subsection{SED Analysis}
\label{sec:sed}

\begin{figure}[]
\includegraphics[scale=0.3, angle=90]{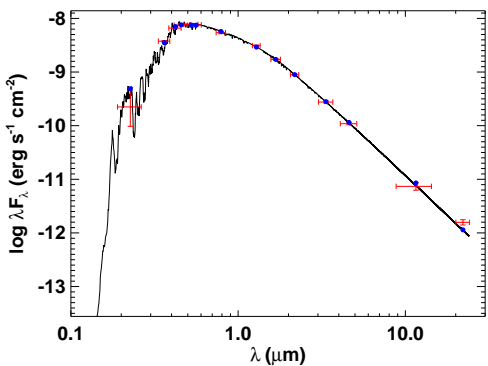}
\caption[]{Measured and best-fit SED for KELT-7 from UV through mid-IR. The red error bars indicate measurements of the
flux of KELT-7 in UV, optical, NIR, and mid-IR passbands and listed in Table \ref{tab:stellar}. The vertical bars are the 
$1\sigma$ photometric uncertainties, whereas the horizontal error bars are the effective widths of the passbands. The 
solid curve is the best-fit theoretical SED from the NextGen models of \citet{hauschildt:1999}, assuming stellar parameters
\teff, \logg, and \feh\ fixed at the adopted values in Table \ref{tab:median}, with $A_{V}$ and $d$ allowed to vary. 
The blue dots are the predicted passband-integrated fluxes of the best-fit theoretical SED corresponding to our observed
photometric bands. The $22$ micron band shows a slight IR excess as discussed in Section \ref{sec:sed}.
\label{fig:sed}
}
\end{figure}

We construct an empirical spectral energy distribution (SED) of KELT-7 shown in Figure \ref{fig:sed}.
We use the near-UV bandpasses from GALEX \citep{martin:2005}, the $B_{T}$ and $V_{T}$ colors from the 
$Tycho-2$ catalog \citep{hog:2000}, near-infrared (NIR) fluxes in the $J$ and $H$ passbands from the 2MASS Point
Source Catalog \citep{cutri:2003, skrutskie:2006}, and near- and mid-infrared fluxes in the four $WISE$ passbands
\citep{wright:2010} to derive the SED. We fit this SED to the NextGen models from \citet{hauschildt:1999} by 
fixing the values of \teff, \logg, and \feh\ inferred from the global fit to the light curve and RV data as 
described in Section \ref{sec:exo} and listed in Table \ref{tab:median}, and then finding the values of the
visual extinction $A_{V}$ and distance $d$ that minimize $\chi^{2}$. We find $A_{V} = 0.13 \pm 0.04$ and 
$d = 129 \pm 8$ pc with the best fit model having a reduced $\chi^{2} = 1.83$. The results from this analysis are
shown in Table \ref{tab:stellar}. We note that the quoted statistical 
uncertainties on $A_{V}$ and $d$ are likely to be underestimated because we have not accounted for the uncertainties
in values of \teff, \logg, and \feh\ used to derive the model SED. Furthermore, it is likely that alternate 
model atmospheres would predict somewhat different SEDs and thus values of extinction and distance. 

Our SED analysis yields a slight IR excess in the $22$ micron band which was also reported by \citet{mcdonald:2012}
in a study of IR excess of Hipparcos stars. Due to the young age of this star (see Section \ref{sec:evo}),  
the detection of this excess could be evidence for a debris disk, though we suspect that it
is likely due to background nebulosity. Inspection of the WISE image in the $22$ micron band shows clear background
nebulosity associated with a nearby bright embedded star forming region that is very bright in teh WISE $22$ micron
image. Therefore we consider it likely that this background nebulosity is the cause of the apparently slight excess
in the WISE $22$ micron passband. In any event, the excess is only $\sim 2 \sigma$ and appears only in this one band, 
therefore it has no impact on the overall SED model fit.

\subsection{UVW Space Motion}
\label{sec:uvw}

We evaluate the motion of KELT-7 through the Galaxy to place it among standard stellar populations. The absolute heliocentric
radial velocity is $+39.4 \pm 0.1$ $ \kms$, where the uncertainty is due to the systematic uncertainties in the absolute 
velocities of the RV standard stars. Combining the absolute TRES RV with the distance from the spectral energy distribution analysis
and proper motion information from 
the UCAC4 catalog \citep{zacharias:2013}, we find that KELT-7 has a $U, V, W$ (where positive $U$ is the direction of
the Galactic center) of $-33.5 \pm 0.2, -9.7 \pm 1.8, -8.4 \pm 0.9$, all in units of \kms, making this a thin disk 
star \citep{bensby:2003}.

\section{False Positive Analysis}
\label{sec:false}

There are many signals that could be mistaken for a planetary transit, so it is important to address some of these false positive
scenarios. There are several reasons to favor a planetary signal over a false positive scenario for KELT-7b.

KELT has a very small aperture, and thus a very large point spread function (PSF), so many initial detections turn out to be blended 
starlight from more than one star 
in the PSF mimicking a transit signal. Therefore, it is important that we follow up our initial detection with seeing-limited telescopes
(i.e., with PSFs of $\sim1^{\prime\prime}$) to rule out any blended eclipsing binaries. Observations using larger telescope in multiple 
filters then resolve stars that are blended even at the $1^{\prime\prime}$ resolution, which typically turn out to be bound systems such as
hierarchical triples.
Our follow-up transits were observed in several different bandpasses ($Vgriz$), and we found no evidence of a wavelength-dependent
transit depth. 

We carefully inspected our spectra to look for light from another source. 
We did not see any evidence for the spectrum being double- or triple-lined. Our bisector analysis of the RVs taken out of transit showed
no indication of being in phase with the orbital solution but we do see a correlation between bisector variation and RV variation of the spectra
taken during transit due to the RM effect.

Our global fit with all spectroscopic and photometric data is well modeled by that of a transiting planet around a single star. 
We find that the \logg\ derived from our global fit, $\loggmulti\pm0.019$, is consistent within 
errors to \logg\ derived from our SPC analysis, $\loggspc\pm0.1$. The amplitude of the RM signal is consistent with that expected from
the \vsini\ measured from the stellar spectrum and the depth and impact parameter measured from the high-precision transit light curves 
(see Section \ref{sec:rm} and Section \ref{sec:exo}).

Finally, we obtained AO images, which exclude companion sources beyond a distance of $0.1^{\prime\prime}$, $0.2^{\prime\prime}$,
$0.5^{\prime\prime}$ and $1.0^{\prime\prime}$ from KELT-7 down to a magnitude difference of $2.5$ mag, $5.4$ mag, $6.4$ mag 
and $7.3$ mag respectively, at a confidence level of $5\sigma$ (see Figure \ref{fig:ao}).

We conclude that all the evidence is best described by a transiting hot Jupiter planet orbiting a rapidly rotating F-star. 
There is no significant evidence suggesting that the signal is better described from blended sources.

\section{Discussion}
\label{sec:discussion}

\begin{figure}[]
\plotone{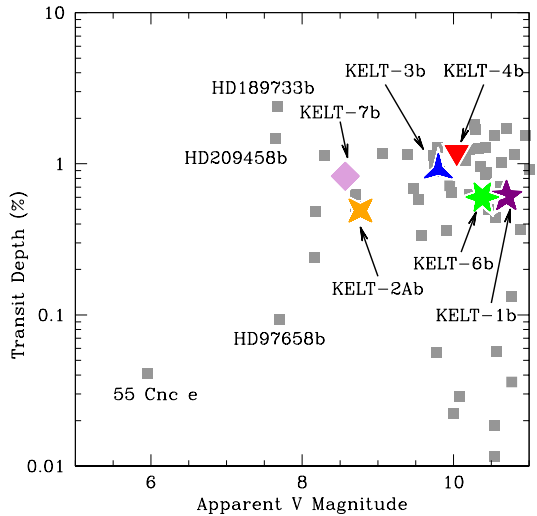}
\caption[]{
Transit depth as a function of apparent V magnitude of the host star for a sample of transiting systems with
bright ($V \leq 11$) hosts. KELT-7b is shown as a pink diamond. Bright stars with deep transits are generally
the best targets for detailed follow-up.
\label{fig:rprs}
}
\end{figure}

We have presented the discovery of KELT-7b, a hot Jupiter planet orbiting the ninth brightest star to host a known transiting planet.
This is the fifth most massive star and fifth hottest star to host a transiting planet\footnote{According to http://exoplanets.org/}.
Figure \ref{fig:rprs} shows the V magnitude versus transit depth for known transiting systems with V$<11$, with the KELT discoveries
highlighted. KELT-7b is an excellent candidate for future detailed atmospheric studies because it is a bright host star
and it has a relatively deep transit. Although we suspect that it may be due to background nebulosity
because the star lies in a region of considerably higher IR nebulosity than its surroundings, the slight IR excess we find
at $22$ microns can be confirmed or excluded using follow up observations.

KELT-7 is a hot ($\sim6800$ K), rapidly-rotating
(\vsini\ $\sim 73$ \kms) star, and its planetary companion was originally
confirmed via the RM effect, which was easily detected with an
amplitude of several hundred \ms.  On the other hand, the reflex
radial velocity motion of the star due to the companion was much more
difficult to detect, although we did ultimately detect the signal at
high confidence and to date this is the second most rapidly rotating transiting system to have this motion measured. 
This discovery therefore illustrates both the
opportunities and challenges associated with confirming 
planetary companions transiting hot stars.  We note that, because of its
brighter magnitude range, the
Transiting Exoplanet Survey Satellite (TESS) \citep{ricker:2014} will also
survey a large number of hot stars.  Therefore, at least some
of the lessons learned from KELT for characterizing the population of planets
orbiting hot stars are likely to apply to TESS as well.

The fast rotation also allowed us to measure the RM effect and measure the projected obliquity of the system.
Understanding the projected spin-orbit alignment of a planet and host star can allow us to infer information
about the formation and evolution of hot Jupiters.
\citet{winn:2010} and \citet{schlaufman:2010}, by different methods, proposed that hot stars ($\teff>6250$ K) that 
host a transiting hot Jupiter typically have high stellar obliquity. \citet{winn:2010} suggested that hot Jupiter 
systems initially have a broad range of obliquities, but the cool stars eventually realign with the orbits of their companions 
because they undergo more rapid tidal dissipation than hot stars. \citet{albrecht:2012} did an RM analysis on 
a sample that nearly doubled the \citet{winn:2010} sample and confirmed the correlation of projected obliquity  
and the effective temperature of the star. \citet{albrecht:2012} also showed that the obliquity of systems  with close-in
massive planets have a dependence on the mass ratio and the distance between the star and planet. Specifically, they
found that higher obliquities are measured in systems where the planet is relatively small. 

The KELT-7 system consists of a transiting hot Jupiter on a fairly close orbit ($a = 0.04$ AU) to its massive and hot host star.
We measured the system to have a low stellar obliquity ($\lambda = 9.7 \pm 5.2$ degrees). One might expect that this planet formed with
a low obliquity and migrated in close to the star because it has been suggested that if the planet formed around a hot host 
star with a high obliquity it would be unable to realign due to the lack of convective envelope. With a larger sample of hot stars with 
transiting planets with projected obliquities, it will become possible to disentangle the dependences of stellar effective temperature,
age, planet mass, and orbital distance on the projected obliquity. Ultimately, this will enable a deeper 
understanding of how systems form and evolve over time, and allow us to distinguish which systems are truly unique.


\acknowledgements

\paragraph{Acknowledgements}
This paper uses observations obtained with facilities of the Las Cumbres Observatory Global Telescope. The Byrne Observatory at Sedgwick
(BOS) is operated by the Las Cumbres Observatory Global Network and is located at the Sedgwick Reserve, as part of the University of 
California Natural Reserve System.

Early work on KELT- North was supported by NASA Grant NNG04GO70G.

A.B. acknowledges partial support from the Kepler mission under Cooperative Agreement NNX13AB58A with the Smithsonian Astrophysical Observatory, D.W.L. PI.

B.J.F. acknowledges that this material is based on work supported by the National Science Foundation Graduate Research Fellowship under Grant No. 2014184874. 
Any opinions, findings, and conclusions or recommendations expressed in this material are those of the authors and do not necessarily 
reflect the views of the National Science Foundation.

R.S.O. acknowledges that this work was performed in part under contract with the California Institute of Technology (Caltech)/Jet Propulsion Laboratory (JPL)
funded by NASA through the Sagan Fellowship Program executed by the NASA Exoplanet Science Institute.

K.G.S. acknowledges support from the Vanderbilt Office of the Provost through the Vanderbilt Initiative in Data-intensive Astrophysics and
the support of the National Science Foundation through PAARE Grant AST-1358862.

Work by B.S.G. and T.G.B. was partially supported by NSF CAREER Grant AST-1056524.

Work by J.N.W. was supported by the NASA Origins program under grant NNX11AG85G.

The authors would like to thank the KELT partners, Mark Manner, Roberto Zambelli, Phillip Reed, Valerio Bozza, that have contributed to the 
project and Iain McDonald for his conversations about the data regarding the IR excess detected in the SED analysis.

This work has made use of NASA Astrophysics Data System, the Exoplanet Orbit Database at exoplanets.org, the Extrasolar Planet Encyclopedia at 
exoplanet.eu (Schneider et al. 2011), the SIMBAD database operated at CDS, Strasbourg, France, and Systemic (Meschiari et al. 2009). 

\end{document}